%% file: paper.tex
\documentclass[acmtog]{acmart}

\renewcommand\footnotetextcopyrightpermission[1]{} 

\usepackage{booktabs}

\input{macros}
\input{macro} 

\newcommand{\revision}[1]{\textcolor{black}{#1}}
\newcommand{\revise}[1]{\textcolor{black}{#1}}

\makeatletter
\newcommand\paragraphNew{\@startsection{paragraph}{4}{\parindent}%
  {-.5\baselineskip \@plus -2\p@ \@minus -.2\p@}%
  {-3.5\p@}%
  {\ACM@NRadjust{\@parfont}}}
	\makeatother
	
\AtBeginDocument{%
  \providecommand\BibTeX{{%
    \normalfont B\kern-0.5em{\scshape i\kern-0.25em b}\kern-0.8em\TeX}}}

\begin{document}

\title{SpongeCake: A Layered Microflake \rev{Surface Appearance} Model}


 \author{Beibei Wang$^{*}$}
\orcid{0000-0001-8943-8364}
\affiliation{%
  \institution{Nankai University and Nanjing University of Science and Technology}
	\streetaddress{College of Computer Science}
  \country{China}}
\email{beibei.wang@nankai.edu.cn}

\author{Wenhua Jin$^{*}$}
\thanks{Beibei Wang and Wenhua Jin contribute equally.}
\affiliation{%
  \institution{School of Computer Science and Engineering, Nanjing University of Science and Technology}
  \country{China}}
\email{jwh@njust.edu.cn}

\author{Milo\v s Ha\v san}
\affiliation{
\institution{Adobe Research}
\country{USA}}
\email{milos.hasan@gmail.com}

\author{Ling-Qi Yan}
\affiliation{
\institution{University of California, Santa Barbara}
\country{USA}}
\email{lingqi@cs.ucsb.edu}

\renewcommand{\shortauthors}{Wang, et al.}

\input{abstract}


\begin{teaserfigure}
  \centering
  \includegraphics[width=\textwidth]{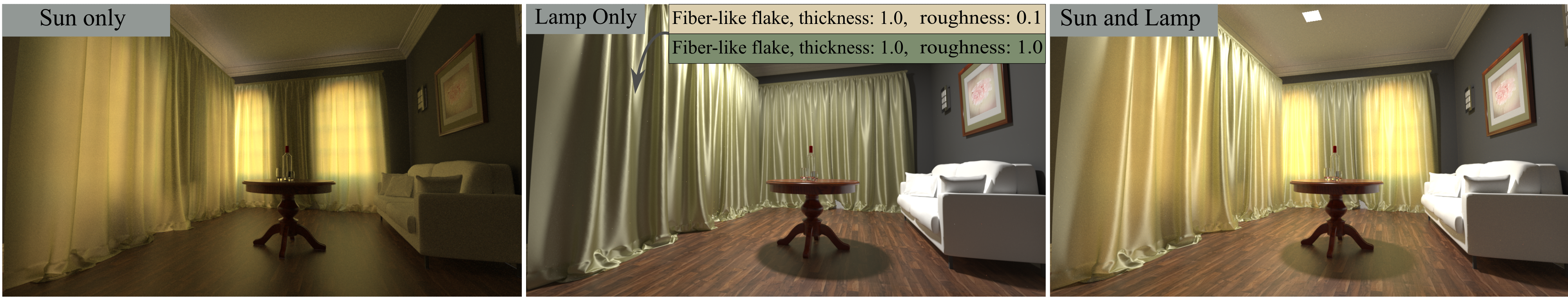}
  \caption{In this indoor scene, we model the window shade using our SpongeCake model, with a two-layer configuration: a specular fiber-like microflake layer on the inside and a rougher fiber-like microflake layer at bottom. Three different light settings are shown: exterior sunlight gives a diffuse (but non-Lambertian) transmission effect, while interior lighting leads to a very different specular fabric sheen effect; finally, we combine the two lighting configurations. Our layered model is able to design these kinds of appearances easily, while offering fast analytic evaluation, including an effective multiple scattering approximation; there is a lack of comparable analytic material models with similar benefits.\label{fig:teaser}}
  \label{teaser}
\end{teaserfigure}

\maketitle

\input{sec_intro}
\input{sec_related}

\input{sec_overview}

\input{sec_single}
\input{sec_multiple}

\input{sec_results}
\input{sec_conclusion}


\bibliographystyle{ACM-Reference-Format}
\bibliography{paper}









\end{document}

%% file: macros.tex
\newcommand{\ra}{\rightarrow}

\newcommand{\dd}{\mathrm{d}}

\newcommand{\rev}[1]{\textcolor{black}{#1}}

\newcommand{\win}{\omega_i}
\newcommand{\wout}{\omega_o}
\newcommand{\wh}{h}
\newcommand{\cti}{\cos \omega_i}
\newcommand{\cto}{\cos \omega_o}

%% file: macro.tex
\def\figurePath{images/}
\def\myfigure#1#2#3{\begin{figure}[tb]\centering\includegraphics[width = \linewidth]{\figurePath#2}\caption{#3}\label{fig:#1}\end{figure}}
\def\mycfigure#1#2#3{\begin{figure*}[tb]\centering\includegraphics*[clip, width = \linewidth]{\figurePath#2}\caption{#3}\label{fig:#1}\end{figure*}}



%% file: abstract.tex
\begin{abstract}

In this paper, we propose SpongeCake: a layered BSDF model where each layer is a volumetric scattering medium, defined using microflake or other phase functions. We omit any reflecting and refracting interfaces between the layers. The first advantage of this formulation is that an exact and analytic solution for single scattering, regardless of the number of volumetric layers, can be derived. We propose to approximate multiple scattering by an additional single-scattering lobe with modified parameters and a Lambertian lobe. We use a parameter mapping neural network to find the parameters of the newly added lobes to closely approximate the multiple scattering effect. Despite the \rev{absence} of layer interfaces, we demonstrate that many common material effects can be achieved with layers of SGGX microflake and other volumes with appropriate parameters.
A normal mapping effect can also be achieved through mapping of microflake orientations, which avoids artifacts common in standard normal maps. Thanks to the analytical formulation, our model is very fast to evaluate and sample. Through various parameter settings, our model is able to handle many types of materials, like plastics, wood, cloth, etc., opening a number of practical applications.


\end{abstract}

%% file: sec_intro.tex
\section{Introduction}
\label{sec:intro}

Rendering layered materials is an important challenge in computer graphics. However, simulating light reflection from a general plane-parallel layer configuration is computationally challenging. Several previous solutions are based on position-free Monte Carlo simulation \cite{Guo:2018:Layered}, which is general and unbiased, but requires expensive path sampling within each BSDF (bidirectional scattering distribution function) evaluation, and introduces additional variance (noise) into the rendering process. Other solutions have been proposed \cite{jakob:2014:layered,belcour:hal-01785457,Zeltner2018Layer,WeierAndBelcour:2020:Anisotropic} but all have significant complexity or other limitations. Instead of fully solving the general layering problem, our goal is to design a more restricted model that is fast, practical, and can represent many common appearances.

In this paper, we propose SpongeCake, a volumetric layered BSDF model. In this model, each layer is a volumetric scattering medium using microflake or other phase functions, and there are no reflecting nor refracting BSDF interfaces between the layers, except for an optional bottom substrate BSDF. In cases where the reflective effects from these interfaces are needed, we show that they can often be approximated by SGGX microflake layers~\cite{Heitz:2015:SGGX} with appropriate parameters.

The absence of top and internal interfaces means that light can pass through the ``fuzzy'' medium boundaries with no change in direction, giving the model its name. This leads to the major benefit of our model: its computational tractability. We derive a general, exact analytic single scattering solution, for both reflection and transmission on any number of volumetric layers, building upon existing single-scattering derivations for more specific cases.

Multiple scattering is important for many materials, and we note the shape of the multiple-scattering lobe is typically similar to the single-scattering lobe, except with modified parameters. The key technical contribution of this paper is an accurate approximation to the multiple scattering effect for the SpongeCake model. We model the multiple scattering component using the sum of two additional lobes, a second analytic single-scattering lobe with modified parameters and a Lambertian lobe. We propose a parameter mapping approach, finding appropriate parameters for the newly added lobes, via a fully-connected neural network. Note that the parameter mapping is performed before rendering, \revision{and is fast even for texture maps, since the network is lightweight}. Our method does not run any neural network inference during rendering, and only needs to handle two extra analytic lobes at render time.

Through various parameter settings, our model is able to handle many types of materials, like plastics, wood, cloth, etc. The parameters of the model can often be related to actual physical parameters, e.g. fiber orientations and colors, such as in \rev{Fig.~\ref{fig:teaser}}. Normal mapping can be achieved in our model through mapping of microflake orientations, which avoids common artifacts of standard \rev{hemispherical} normal mapping. Thanks to the analytical formulation, our model is very fast to evaluate and sample, and does not introduce additional noise (variance).
In summary, the contributions of our paper are:
\begin{itemize}
    \item Definition of the SpongeCake model, and demonstration of the wide range of appearances it can achieve: plastics, metals, fabrics, plant leaves, and polished hardwood with secondary specular highlights.
    \item An analytic single-scattering solution, including reflection and transmission of multiple layers, together with an importance sampling scheme. This is a generalization of many similar previous single-scattering derivations.
    \item An accurate analytic approximation to multiple scattering, based on the same analytic form as the single-scattering BSDF, with modified parameters predicted by a neural network.
\end{itemize}

The SpongeCake model can be used, with or without multiple scattering, to describe a broad set of material appearances that would be more expensive or less convenient to represent with alternative methods. We believe it can become a standard tool in the physically-based shading toolkit.

%% file: sec_related.tex
\section{Related Work}
\label{sec:related}

%
\emph{Layered material models}
are the closest to our SpongeCake model. The earliest in computer graphics is the model by Hanrahan and Krueger~\shortcite{hanrahan1993reflection}, where an analytic single scattering model is derived. This is similar to our solution, but less general, and uses Monte Carlo simulation to consider multiple scattering. Later, three different branches of layered material research were proposed. The first line of research uses discretized BSDFs for layering. A representative work was by Jakob et al.~\shortcite{jakob:2014:layered} that uses Fourier-space discretized BSDFs to do layering, which was later extended by Zeltner and Jakob~\shortcite{Zeltner2018Layer} to support anisotropy. Another direction in layered materials is analytic models. Weidlich and Wilkie~\shortcite{weidlich2007arbitrarily} derived an analytic layered model from individual multi-lobe BSDFs, though with various approximations and not considering multiple scattering. Belcour~\shortcite{belcour:hal-01785457} uses low-order moments of BSDFs and achieves fast performance, but cannot support anisotropy nor arbitrary volumetric layers. Further improvements were introduced by more recent work of Weier and Belcour \shortcite{ WeierAndBelcour:2020:Anisotropic} and Yamaguchi et al.~\shortcite{Yamaguchi:2019:anisotropic}. These methods extend the capabilities of previous solutions towards anisotropy and other effects, but at the expense of more involved mathematical techniques. In contrast, our work restricts the model itself, such that exact single scattering and accurate approximate multiple scattering can be achieved, while keeping the model relatively simple. Furthermore, Belcour's and the follow-up works can handle neither fiber-like appearance (e.g., wood, fabric and sheen effects), nor volumetric media.

Another line of work was initiated by Guo et al.~\shortcite{Guo:2018:Layered}, using a position-free path integral formulation that works with general layered materials. This approach applies Monte Carlo estimators to the path integral to achieve efficient direct illumination (BSDF evaluation). This work was later improved by Xia et al.~\shortcite{Xia:2020:Layered} and Gamboa et al.~\shortcite{Gamboa:2020:EfficientLayered} with more advanced sampling approaches and more efficient estimators.  Guill\'{e}n et al.~\shortcite{Guillen:2020:Pearlescent} leverage the generality of position-free MC for modeling pearlescent pigments. Though fairly efficient and versatile, these methods produce Monte Carlo noise (variance), and still cannot compete with analytic solutions in terms of performance. They are also unlikely to be usable in real-time applications. In contrast, our model is more restricted, but is entirely analytic at rendering time, leading to very fast evaluation and no additional Monte Carlo noise.

Note that most layered material models assume reflective and refractive interfaces between different layers, due to their differing indices of refraction. These interfaces are usually modeled as either perfectly specular or microfacet BSDFs~\cite{Stam2001}\cite{Walter07}. In our work, we omit such interfaces (except for the optional bottom substrate BSDF), instead treating all layers as volumetric media made of scattering particles in air. This greatly simplifies our model, leading to an analytic solution for single scattering and an approximate analytic multiple scattering, and we show how to recover the reflective effects of the interfaces by well-chosen microflake media in our results. On the other hand, our model is not capable of refraction, so it cannot model e.g. rough glass solids (though it could still be used to layer additional effects on top of a refractive BSDF). We find this to be an acceptable trade-off, since in return we get fast and accurate evaluation of single and multiple scattering, and we still support a large range of appearances.


\paragraph{Microflake models.} Our SpongeCake model defines each layer as a volumetric medium, typically (though not necessarily) defined by a microflake distribution. The concept of microflake distributions and phase functions was introduced by Jakob et al~\shortcite{Jakob:2010:microflake} to accurately define the full anisotropy of participating media, by modeling the distribution of underlying basic reflective elements called microflakes. The microflake model is successful in representing fibers and fabrics~\cite{Zhao:2011:fabric}. Other work has also benefited from using microflakes for representing other types of materials such as
foliage~\cite{LoubetNeyret:2018:microflake}, or special pigments~\cite{Guillen:2020:Pearlescent}.

Heitz et al.~\shortcite{Heitz:2015:SGGX} introduced a versatile representation named symmetric GGX (SGGX) to efficiently describe microflake distributions using $3 \times$ 3 positive-definite matrices, allowing for surface-like and fiber-like microflakes, including convenient control over their orientations. We use this formulation in many of our results, though our framework is not restricted to it. We currently assume symmetric microflakes that act as mirrors on both sides.

\paragraph{Microfacet multiple scattering.} Heitz et al.~\shortcite{heitz2016} also introduced a multiple-scattering model for microfacet BSDFs, which is based on designing a microflake volume in a very specific manner, so that its single scattering exactly matches the behavior of microfacet models. Once such a volume is defined, multiple scattering in the microgeometry can be simulated by Monte Carlo random walks. Matching microfacet models with microflake volumes requires single-sided microflakes and other modifications. 
This research is in a fairly different direction from ours. In our model, we accept that uniform layers of two-sided mirror microflakes do not exactly match microfacet models. We focus on analytic evaluation, avoiding Monte Carlo computations. Adding multiple scattering to our model is a distinct problem from adding it to microfacet models, and addressing the latter is not in our scope. \revise{Still, these research directions may become more unified in the future.}

Another line of work approximated the multiple scattering, by introducing a separable correction term \cite{Ashikmin:2000:Microfacet,Kelemen01amicrofacet,jakob:2014:layered,KullaConty:2017:revisiting} or by scaling the singe scattering lobe \cite{Turquin:2019:multiple}. These approaches are efficient and suitable for practical rendering systems, but the shapes of their BSDF lobes do not match simulations of actual microfacets, and in Turquin's case also lose the reciprocity property.

Xie et al.~\shortcite{Xie:2019:multiple} proposed to represent the multiple scattering with the Real NVP neural architecture, and used these models for rendering at run-time. Different from their work, our neural network only predicts parameters of additional lobes as a preprocess, while at rendering time it allows for analytic evaluation and sampling. Their work also does not consider microflakes nor layering.

All the multiple-scattering models mentioned above use the Smith shadowing model. Another group of works~\cite{lEE2018PRATMUL, Feng2018multiV} use the V-groove model, which allows for analytic solutions for multiple scattering. However, they suffer from several issues: overly shiny results for rough surfaces, discontinuous derivatives, or lack of compatibility with dielectric materials.

\paragraph{Single scattering in volumetric layers.} Our method is inspired by a series of previous work that derive single scattering from volumetric configurations in multiple forms. Chandrasekhar~\shortcite{Chandrasekhar:1960} is a standard reference for radiative transport, and describes a BRDF for half-space isotropic scattering. Later, this BRDF was extended for lunar regoliths \cite{Hapke:1963:lunar,Hapke:1981:spectroscopy} and for subsurface scattering \cite{Premoze:2002:Volumetric}. Koenderink and Pont~\shortcite{KoenderinkPont:2003:skin} proposed an asperity scattering lobe, which also considers single-scattering in an isotropic layer. Recently, d'Eon~\shortcite{dEon:2021:analyticLambertian} proposed a new analytic BRDF for porous materials where scattering and absorption is approximated by spherical Lambertian particles (rather than an isotropic phase function, which is a different model).

Dupuy et al.~\shortcite{Dupuy:2016:unified} further explored the deep connections of the microflake theory in volumes with the microfacet theory on surfaces. Their framework derives the BRDF of a semi-infinite slab of microflake media, but because they consider a specialized asymmetric version of microflake theory, designed to match microfacet theory, their result is not equivalent to ours.

The main difference from previous work is that we give the result in a form specific to standard (two-sided) microflakes, while previous results either precede microflake theory or use an uncommon modification of it. Our single-scattering solution also works for any number of layers and for both reflection and transmission.


\paragraph{Neural networks for material appearance} Neural networks are commonly used in rendering, to replace various parts of the rendering process. Here we cover only the work most closely related to our application (material appearance and multiple scattering). Kallweit et al.~\shortcite{kallweit2017deep} used a moderately small neural network to predict multiple scattered energy between two different places in the volume of a cloud. Yan et al.~\shortcite{yan2017furbssrdf} proposed a lightweight multi-layer perceptron (MLP) to convert fur fiber properties to participating media scattering properties. Kuznetsov et al.~\shortcite{Kuznetsov19:GlintsGAN} trained a Generative Adversarial Network (GAN) to dynamically create new Normal Distribution Functions (NDFs) to render detailed appearance. Rainer et al. \shortcite{Rainer2019Neural,Rainer2020Unified} compress measured appearance data using autoencoders; the neural decoder needs to be evaluated during rendering. Recently, neural networks have also been used for importance sampling of BRDFs \cite{sztrajman2021nbrdf} and BSSRDFs (bidirectional surface scattering distribution functions) \cite{ Vicini:2019:LearnedShape,Leonard:021:Volumetric}. We design a lightweight neural network to predict a set of single scattering parameters that would render similar to a given multiple scattering appearance; this is done as a preprocess to actual rendering, during which our model's evaluation is analytic.

%% file: sec_overview.tex
\input{paper_table}

\section{Background and Overview}
\label{sec:overview}

In this section, we briefly review some background knowledge, and introduce our model at a high level. SpongeCake is a layered BSDF model, where each plane-parallel layer contains a homogeneous volumetric medium. The volumetric media within the layers can have any absorption and scattering properties and can use various phase functions, including microflake-based, but also more traditional ones like Henyey-Greenstein. Therefore, we start with a general look at microflakes and their distributions.


\subsection{General definitions}

Our notation is summarized in Table \ref{tab:notations}. We will denote incoming (light) direction by $\omega_i$ and outgoing (camera) direction by $\omega_o$. Consider a homogeneous volumetric layer of thickness $T$. The volumetric medium has a phase function $f_p(\win \ra \wout)$, and a directionally-dependent extinction coefficient $\sigma_t(\omega)$. For microflake phase functions, the extinction coefficient depends on the direction $\omega$, while for other phase functions it is typically constant. The extinction coefficient should be symmetric: $\sigma_t(\omega) = \sigma_t(-\omega)$. The phase functions should satisfy reciprocity and energy conservation respectively (\cite{Jakob:2010:microflake}):
\begin{equation} \label{eq:recip}
    \sigma_t(\win) f_p(\win \ra \wout) = \sigma_t(\wout) f_p(\wout \ra \win),
\end{equation}
and
\begin{equation} \label{eq:energy}
    \int_{S^2} f_p(\win \ra \wout) \dd \wout = 1.
\end{equation}
For the exposition of our model, the generalized extinction coefficient $\sigma_t$ is sufficient. We could also consider generalized absorption and scattering coefficients, but they are not necessary for the exposition below, and we omit them for simplicity.

\subsection{Microflake media}

For our purpose, microflake media can be thought of as volumetric distributions of tiny, flat two-sided mirrors, called microflakes (though one-sided and diffuse microflakes can also be defined). Each flake can have a constant or a directionally-varying reflectance (analogous to a Fresnel term of the mirror). The normals of the microflakes follow a directional distribution $D: \mathcal{S}^2 \rightarrow \mathcal{R}^+$. 

Next, we define the projected area of the microflakes as:
\begin{equation}
    \sigma(\omega) = \int_{S^2} D(m) \langle \omega, m \rangle \dd m,
\end{equation}
where $\langle x, y \rangle$ is the clamped (non-negative) dot product. Note the lack of subscript on $\sigma$. The extinction function $\sigma_t$ is defined to be the projected area, scaled by the microflake density $\rho$: $\sigma_t(\omega) = \rho \sigma(\omega)$. The units of $\sigma_t$ and $\rho$ are mm$^{-1}$, while the quantity $\sigma(\omega)$ is unitless.

We will assume that $D(\omega)$ is symmetric, and the microflakes are double-sided mirrors; then the microflake phase function can be written as (\cite{Heitz:2015:SGGX}):
\begin{equation} \label{eq:phase}
f_p(\win \ra \wout) = \frac{D(\wh)}{4 \sigma(\win)},
\end{equation}
where $\wh$ is the half vector between $\win$ and $\wout$. This definition satisfies reciprocity and energy conservation. It is not necessary that $D$ is normalized as a pdf; though this could be achieved if desired, as scaling $D$ by a constant will scale $\sigma$ by the same constant, canceling out in Eqn.~(\ref{eq:phase}).



For convenience of our future derivation, we define a \emph{reduced phase function} as
\begin{equation}
\hat f_p(\win \ra \wout) = \sigma_t(\win) F(\wh) f_p(\win \ra \wout),
\end{equation}
where $F(\wh)$ is a reflectance term, which could be either constant or represent a Fresnel-like term. The reduced phase function appears at each scattering vertex of a light path through the volumetric layer. The reduced phase function is also reciprocal in the usual sense:
\begin{equation} \label{eq:recip2}
    \hat f_p(\win \ra \wout) = \hat f_p(\wout \ra \win).
\end{equation}


\subsection{SGGX microflakes}

While the above can work with any microflake distribution, in our results we use the SGGX microflake \cite{Heitz:2015:SGGX} with parameters designed to achieve either a surface-like or a fiber-like distribution. Let $\alpha$ be a roughness parameter, analogous to the roughness in the GGX NDF \cite{Walter07}. The SGGX distribution is controlled by a symmetric, positive definite $3 \times 3$ matrix $S$ (defined below), and can be written as:
\begin{equation}
    D(\omega) = \frac{1}{\pi \alpha q^2}, \mbox{ where }
    q = \omega^\top S^{-1} \omega, \mbox{ and } \sigma(\omega) = \sqrt{\omega^\top S \omega}.
\end{equation}
Note that the expression $q = \omega^\top S \omega$ is a quadratic form, and the surface where $q=1$ is an ellipsoid \revise{(if allowing for unnormalized $\omega$)}. This leads to the intuition of modeling fibers as long, skinny ellipsoids, and surfaces as flattened ellipsoids. The ellipsoids can be rotated to model fiber directions or normal maps.

The $3 \times 3$ matrix $S$ specifying the SGGX distribution is $\mbox{diag}(1,1,\alpha^2)$ for fiber-like distributions, and $\mbox{diag}(\alpha^2,\alpha^2,1)$ for surface-like distributions. We can rotate these axis-aligned definitions using any $3 \times 3$ rotation matrix $R$, obtaining a rotated SGGX matrix $S' = R^\top S R$.

Our approach works with any analytic volumetric phase functions; in most cases, we will use microflake phase functions, for which SGGX microflakes are a specific convenient definition.

\subsection{SpongeCake overview}
Since our SpongeCake model consists of only volumetric layers, all capabilities of this model can be evaluated with Monte Carlo random walk simulation of light transport, which we use as ground truth. However, our goal is to achieve much more efficient analytic evaluation and sampling. 

We first derive an analytic single scattering formula for our BSDF model (Sec.~\ref{sec:single}). We show that for a microflake phase function with flake distribution $D$, the resulting BSDF has a form similar (but not identical) to a standard microfacet BSDF \cite{Walter07}, with $D$ becoming analogous to the microfacet normal distribution function. We discuss extensions to transmission and multiple layers. We also discuss the optional unscattered delta transmission component, which we may or may not choose to include, depending on the desired appearance.

While single scattering is already sufficient for a variety of appearances, multiple scattering is important for some materials. We approximate multiple scattering by noting that the ground truth multiple-scattering component of the BSDF often looks similar to the single-scattering component, except with modified parameters; an observation already made by others \cite{heitz2016}. We propose a multiple-to-single mapping approach, finding the parameters via a small neural network (Sec.~\ref{sec:multiple}). Thus, the multiple scattering term has the same form and the same evaluation cost as the single-scattering term.

%% file: paper_table.tex
\begin{table}[!t]
	\renewcommand{\arraystretch}{1.1}
	\caption{\label{tab:notations} Notations.}
\begin{small}
  \begin{tabular}{|l|c|l|}\hline
		$\omega_i $ & incident direction \\
		$\omega_o $ & outgoing direction \\
		$h $ & half vector of $\omega_i$ and $\omega_o$ \\
		$f_p(\omega_i \rightarrow \omega_o) $ &  phase function \\
		$\hat{f}_p(\omega_i \rightarrow \omega_o) $ &reduced phase function \\
		$\sigma_t(\omega)  $ &generalized extinction coefficient\\
		$D(\omega)$ &directional distribution function \\
		$\sigma(\omega)  $ &projected area of microflakes\\
		$\rho             $   & microflake density\\
		$F(h)$ &reflectance term \\
		$S$  &3 × 3 matrix for microflake distribution \\
		$\alpha$ & roughness \\
		$T$ & thickness (different from $\top$) \\
		$\top$ & matrix transpose (different from $T$) \\
		$f_r(\omega_i, \omega_o)$ & BRDF \\
		$G(\omega,\omega_m)$ &shadowing-masking term \\
		$\tau_k(\omega)$ & attenuation\\
				\hline
		\end{tabular}

\end{small}
\end{table}

%% file: sec_single.tex
\section{Single scattering}
\label{sec:single}

\myfigure{ss_reflection}{ss_reflection.pdf}{Illustration of BRDF: the single scattering contribution of a layer can be seen as an integral over the depth of the single scattering vertex. The incident direction and the outgoing direction are on the same side of the layer. }


We will first consider a single scattering event in a single reflective layer, under any general phase function. Afterwards, we will specifically consider microflake phase functions, and extend the discussion to transmission and multiple layers.

\subsection{Single-scattering layer derivation}
Consider a homogeneous microflake layer of thickness $T$. As explained by the position-free Monte Carlo framework~\cite{Guo:2018:Layered}, the BRDF of the single scattering contribution of any layer can be written as an integral over the depth of the single scattering vertex. A reduced phase function term corresponds to the scattering vertex, and volume extinction terms as well as reciprocal cosine terms occur on both the incoming and outgoing segments. See also Fig.~\ref{fig:ss_reflection}.
\begin{align}
\label{eq:bsdf1}
    f_r(\win, \wout) = \int_0^{\rev{T}} & \frac{\hat f_p(\win \ra \wout)}{\cti \cdot \cto} \cdot \nonumber \\
    & \exp \left(- \frac{t \sigma_t(\win)}{\cti} \right)
    \exp \left(- \frac{t \sigma_t(\wout)}{\cto} \right)
     \dd t.
\end{align}
Here, we use $\cos \omega$ to mean the cosine of the angle between $\omega$ and the macro surface normal, i.e. $(0,0,1)^\top$ in the local shading frame, which is equal to the $z$-component of $\omega$. The same result can also be derived from first principles using the radiative transfer equation, by considering a ray into the medium from direction $\wout$, under a directional light from direction $\win$ with unit surface irradiance. The cosine terms in the denominator originate from the change of integration domain from incoming ray length to layer depth ($\cos \wout$), and from the definition of a BRDF as outgoing radiance per unit incoming irradiance ($\cos \win$).  

To simplify the integral, we note that only the exponential extinction terms depend on the integration variable; the other terms can be taken out of the integral. Define $\Lambda(\omega) = \sigma(\omega) / \cos \omega$ for brevity. The $\Lambda$ function here is analogous to the one used in previous work on Smith microfacet shadowing/masking, and has been used in the context of microflakes \cite{Dupuy:2016:unified}. This lets us simplify the integral as follows:
\begin{equation}
\label{eq:bsdf2}
    f_r(\win, \wout) = \frac{\hat f_p(\win \ra \wout)}{\cti \cdot \cto}
    \int_0^{\rev{T}} e^{-t \rho (\Lambda(\win) + \Lambda(\wout))}
     \dd t.
\end{equation}
This integral is simple to solve, and the result is:
\begin{equation}
\label{eq:bsdf}
    f_r(\win, \wout) = \frac{\hat f_p(\win \ra \wout)}{\cti \cdot \cto} \cdot
    \frac{1 - e^{-T \rho (\Lambda(\win) + \Lambda(\wout))}}{\rho (\Lambda(\win) + \Lambda(\wout))}.
\end{equation}
%

The above result is valid for any phase function. For the important case of a microflake phase function, we can expand the definition of $\hat f_p$ and arrive at the following BRDF form:
\begin{equation} \label{eq:bsdf-flake}
    f_r(\win, \wout) = \frac{F(h) \ D(h) \ G(\win,\wout)} {4 \cti \cdot \cto},
\end{equation}
where
\begin{equation} \label{eq:gterm}
    G(\win,\wout) = \frac{1 - e^{-T \rho (\Lambda(\win) + \Lambda(\wout))}}{\Lambda(\win) + \Lambda(\wout)}.
\end{equation}

We immediately note the similarity of the formula to a standard microfacet BRDF, though with a different ``shadowing-masking term'' $G$. The model only depends on the product $T\rho$, rather than thickness and density separately. The product of these quantities is the unitless ``optical depth''. It expresses the intuition that increasing the thickness and density of a layer by the same factor has the same effect on appearance. Note also that the resulting BRDF is reciprocal.

\subsection{Comparison to previous derivations}

The above result is similar to previous derivations of single scattering in volumetric slabs of finite or infinite depth. The main difference is that we are the first to explicitly give the result for standard (symmetric) microflake distributions, while previous work either preceded the microflake concept \cite{Chandrasekhar:1960,hanrahan1993reflection} or considered a special asymmetric modification of microflake theory designed to match microfacet theory, which has different behavior \cite{Dupuy:2016:unified}.

Let us first consider the well-known single scattering BRDF for an isotropic semi-infinite slab, as given by Chandrasekhar \shortcite{Chandrasekhar:1960}:
\begin{equation} \label{eq:chan}
    f_r(\win, \wout) = \frac{a}{4 \pi} \frac{1}{\cti + \cto}.
\end{equation}
This result is a special case of our Eqs.~(\ref{eq:bsdf-flake}) and (\ref{eq:gterm}). Note that isotropic scattering is a special case of SGGX scattering for $\alpha = 1$, leading to $D(h) = 1/\pi$ and $\sigma(\omega) = 1$. We further set $F(h) = a$ and $T \ra \infty$ in the limit. The result simplifies to Eqn. (\ref{eq:chan}), confirming our derivation for a special case.

For a finite thickness, the derivation by Hanrahan and Krueger \shortcite{hanrahan1993reflection} is also a special case of ours, see their equation for back-scattered radiance on page 169 (bottom left).

Finally, Dupuy et al. \shortcite{Dupuy:2016:unified} derive the solution for a special version of microflake theory, constructed to match microfacet theory. Their result is different from ours specifically in the $G$ term, where their solution leads to the bistatic (height-correlated) Smith shadowing-masking term:
\begin{equation}
    G(\win,\wout) = \frac{1}{1 + \Lambda(\win) + \Lambda(\wout)},
\end{equation}
which differs from our Eqn.~(\ref{eq:gterm}) for $T \ra \infty$ by the addition of 1 in the denominator. The definitions of $\Lambda$ and $\sigma$ functions are also different due to the different microflake model.

Note that we are interested in the integral of attenuations for all depths in Eqn. (\ref{eq:bsdf2}). However, the derivation technique used by Dupuy et al. (Eqs. (20) and (21)) is to take the expectation over depths sampled from an exponential distribution. This is not equivalent to our integral, and works in their specially constructed framework but not in our setting. Therefore, our results are not a special case of theirs, nor are theirs a special case of ours.

\subsection{Transmission}

\myfigure{ss_transmission}{ss_transmission.pdf}{Illustration of bidirectional transmittance distribution function (BTDF). The incident direction and the outgoing direction are at different sides of the layer.}

Similar to the reflective BRDF, the transmissive BTDF of the single scattering contribution of this layer can be written as an integral over the depth of the single scattering vertex, as shown in Fig.~\ref{fig:ss_transmission}. Assume $\win$ is above and $\wout$ below the horizon (i.e. $\cos \wout < 0$), then
\begin{multline}
\label{eq:bsdf3}
    f_r(\win, \wout) = \int_0^{\rev{T}} \frac{\hat f_p(\win \ra \wout)}{|\cti| \cdot |\cto|}  \\
    \exp \left(- \frac{t \sigma_t(\win)}{\cti} \right)
    \exp \left(- \frac{(t-T) \sigma_t(\wout)}{\cto} \right)
     \dd t.
\end{multline}

Note that we have replaced $(T - t) / |\cos \wout|$ by $(t - T) / \cos \wout$, since we know that $\cos \wout < 0$.
This will result in the same formulation as above, except the $G$ term will be scaled by an additional constant term:
\begin{equation}
    G(\win,\wout) = \frac{1 - e^{-T \rho (\Lambda(\win) + \Lambda(\wout))}}{\Lambda(\win) + \Lambda(\wout)} \cdot e^{T\rho \Lambda(\wout)}.
\end{equation}

Note that we still define $\Lambda(\omega) = \sigma(\omega) / \cos \omega$, so the value of $\Lambda$ will be negative for directions under the horizon. 

\subsection{Single scattering for multiple layers}
\label{sec:layeredsingle}

\myfigure{ss_multilayer}{ss_multilayer.pdf}{Illustration of single scattering for multiple layers. The result becomes a sum over several cases, each corresponding to a scattering event in the respective layer.}

Thanks to the absence of the top and internal interfaces in our SpongeCake model, the above derivation can be trivially extended to a single-scattering model for multi-layer materials, since the light passing through the medium boundaries will not change its direction. Therefore, the integration from $0$ to $T$ in Eqn.~(\ref{eq:bsdf1}) can be naturally segmented into separate layers, and we extend single scattering to multiple layers by summing up the contribution from each layer.
For each layer $k$, we also need to take into account the attenuation of light going through all the layers above the $k$-th in direction $\omega$ (or below, for directions $\omega$ under the horizon), as shown in Fig.~\ref{fig:ss_multilayer}.



\subsection{Delta transmission term and substrate BSDF}
\label{sec:delta}

For transmissive BTDFs, an unscattered (delta function) component is also present when $\win = -\wout$, and is simply the product of transmissions through all layers in the direction $\win$. We can decide to include this component or not, depending on circumstances. Some materials have this effect (e.g. thin fabrics where some rays can travel through without hitting fibers) while others do not (thicker fabrics, leaves, rough refractive plastic, etc.) Inclusion of this component is thus controlled by a boolean flag. We currently ignore this delta transmission term in all our test scenes except in the white furnace test. 

Finally, we can choose to stack volumetric layers on top of a single \emph{substrate} BSDF, which could use any standard model (Lambertian, microfacet, etc.), and could use refraction. The value of this BSDF will again be reduced by transmission terms, for both $\win$ and $\wout$.

\subsection{Importance sampling}
\label{sec:layeredSample}


Importance sampling is an important operation for a BSDF. Given a direction $\win$, for a single layer, we sample the outgoing direction $\wout$ by importance sampling the phase function of the microflake layer. We conduct perfect importance sampling according to the shape of the phase function. Since the phase function itself is normalized by definition, the pdf value is exactly the phase function value.

To sample our multi-layered model, given an incoming direction $\win$, we first obtain a layer $k$, by importance sampling the attenuation along the layers using a discrete probability distribution. If the delta transmission component is desired, we can also choose it here with an appropriate probability.

Once the layer is chosen, we sample the $\wout$ by importance sampling the phase function at layer $k$. With the sampled $\wout$, we evaluate the BSDF value and compute its pdf. The pdf is computed by summing up the pdfs at each layer, which is the phase function value, weighted by layer probabilities for normalization. The sampling weight is then simply the BSDF evaluation divided by the pdf.

Note that we only use single scattering for importance sampling, even after introducing the multiple scattering term, for simplicity. Without the choice between single and multiple scattering, less variance is produced.


%% file: sec_multiple.tex
\myfigure{netSingle}{netsinglelayer.pdf}{The neural network structure for our single layer multiple scattering. The unmodified parameter means that the value of current output parameter is set as the corresponding input parameter, and will not be changed during training. In the current network, the flake orientation $\omega_p$ and phase function type $k$ are unmodified parameters.}

\myfigure{netTwoLayer}{nettwolayer.pdf}{The neural network structure for our two-layer multiple scattering. In the current network, the unmodified parameters for the top and bottom layers are the flake orientation $\omega_p$ and phase function type $k$.} 


\mycfigure{MIS}{MIS.pdf}{Our analytic single scattering model with different sampling approaches: BSDF sampling, light sampling and their combination using multiple importance sampling (MIS). Monte Carlo simulation of the single scattering is used as the ground truth. This serves to validate the correctness of our single scattering model. Since our model supports MIS, it can be added to any common path tracing frameworks. We use a two-layer model: the top layer is a surface-like SGGX model with roughness 0.5, thickness as 0.5, and the bottom map is a surface-like SGGX model with roughness 0.9, thickness as 2.}

\mycfigure{singleLayerMultiple}{singlelayervalidation.pdf}{Single + multiple scattering validation for a set of single-layer materials. For each example, we list the microflake type, roughness $\alpha$ and thickness $T$. The ground truth is generated using Monte Carlo simulation of both single scattering and multiple scattering. We find small differences from the ground truth; see also Fig.~\ref{fig:singleLayerLobe} for the corresponding lobe visualizations.}

\mycfigure{singleLayerMultiple_only}{singleLayerMultiple_1.jpg}{Multiple scattering validation for a set of single-layer materials, over varying thickness $T$ and roughness $\alpha$. Monte Carlo simulation of multiple scattering is used as the ground truth. We find small differences from the ground truth. More examples are provided in the supplemental material.}

\mycfigure{singleLayerLobe}{singlelayervalidationLobe.jpg}{Lobe visualizations for multiple scattering for a set of single-layer materials. Monte Carlo simulation of the multiple scattering is used as the ground truth. For each example, we list the microflake type, reflectance $\gamma$, roughness $\alpha$ and thickness $T$. The first two columns represent the entire BSDF (top and bottom hemispheres), with pixel rows corresponding to a discretization of incoming directions, and pixel columns corresponding to outgoing directions. The latter four columns visualize the outgoing lobe given a fixed incoming direction at the specified angle  $\theta$ {(in radians)} with pixel rows corresponding to the elevation angle of the outgoing direction, and pixel columns corresponding to the azimuth angle of the outgoing directions.}

\mycfigure{twoLayerMultiple}{twolayerValidation.png}{Single + multiple scattering validation for a set of two-layer materials. Monte Carlo simulation of both single and multiple scattering is used as the ground truth. For each example, we list the microflake type, roughness $\alpha$ and thickness $T$ of both layers. See also Fig.~\ref{fig:twoLayerMultiple} for the corresponding lobe visualizations.}

\mycfigure{twoLayerMultiple_only}{twolayerMultiple_1.jpg}{Multiple scattering validation for a set of two-layer materials, over varying thickness and roughness. \revise{The modified parameters are from the bottom layer, while the uppermost layer remains fixed.} Monte Carlo simulation of multiple scattering is used as the ground truth. The differences from the ground truth tend to be small. More examples are shown in the supplemental materials.}

\mycfigure{twoLayerLobe}{twolayerValidationLobe.jpg}{Lobe visualization for multiple scattering validation for a set of two-layer materials. Monte Carlo simulation of the multiple scattering is used as the ground truth. For each example, we list the microflake type, reflectance $\gamma$, roughness $\alpha$ and thickness $T$ of each layer. The first two columns represent the entire BSDF (top and bottom hemispheres), with pixel rows corresponding to a discretization of incoming directions, and pixel columns corresponding to outgoing directions. The latter four columns visualize the outgoing lobe given a fixed incoming direction at the specified angle $\theta$ (in radians) with pixel rows corresponding to the elevation angle of the outgoing direction, and pixel columns corresponding to the azimuth angle of the outgoing directions.}

\myfigure{compareMicrofacet}{compareMicrofacet.pdf}{Comparison between our model \revise{(single scattering only)} using a specular layer of SGGX microflakes and the microfacet model with varying roughness. We scale our phase function by $2\times$, to match the reflectance defined using SGGX (two-sided, phase function normalized on the entire sphere) and GGX (one-sided, normal distribution normalized on the projected hemisphere). We set the thickness of our layer to infinity. We also show the difference images.}


\section{Multiple scattering}
\label{sec:multiple}

When volumetric layers have a large scattering albedo and/or a large thickness, multiple scattering can become significant. However, an analytical solution for multiple scattering in a microflake layer, or multiple such layers, appears to be intractable. A Monte Carlo random walk with next event estimation \cite{Guo:2018:Layered} is a valid way to compute the multiple scattering, however it is costly and adds variance to the rendered result.

Previous work has observed that multiple scattering produces a distribution similar in shape to single scattering, which was used by Turquin~\shortcite{Turquin:2019:multiple} to approximate multiple scattering simply as a rescaled single scattering lobe. However, we find that this simple rescaling is not as accurate, and also loses reciprocity. We propose to instead add another single-scattering lobe with modified parameters, as well as a Lambertian term to more closely approximate the multiple scattering BSDF component. Thus, our full BSDF with multiple scattering is computed as a sum of three lobes: the original (exact) single-scattering lobe, a second single-scattering lobe with modified parameters, and a Lambertian lobe.

Once the parameters of the two newly added lobes are found, evaluation of approximate multiple scattering becomes as simple as computing another analytic single-scattering BSDF with the same form, same number and types of layers, but different parameters (plus a Lambertian term of negligible cost). The remaining challenge is how to obtain the new lobe parameters.

We propose a neural network model to map the original layer parameters to the new lobe parameters. We train the network using a differentiable implementation of our analytic single scattering model. The ground-truth multiple scattering component is computed by Monte Carlo simulation.

The structure of the neural network is a multi-layer perceptron (MLP), consisting of three internal layers, each containing 128 neurons, as shown in Fig.~\ref{fig:netSingle}. We train separate networks for different numbers of layers. We illustrate this for one and two layers; more layers could be added in the same way as desired.

The input to the network is the concatenation of parameter vectors of each layer, including: roughness $\alpha$, single-scattering albedo $\gamma$ (three channels), thickness $T$, Schlick reflectance $f_0$ (three channels), the phase function type $k$ (fiber or surface microflake) and orientation $\omega_p$ (three channels). The output are the parameters for equivalent single scattering, including: roughness $\alpha$, albedo $\gamma$ (three channels), thickness $T$, $f_0$ (three channels), the phase function type $k$ (fiber or surface microflake), orientation $\omega_p$ (three channels) and the weights $W_1$ and $W_2$ for the single scattering term and Lambertian term, where the phase function types are the same as the input.

\paragraph{Dataset} We randomly sample roughness, albedo, reflectance ($f_0$), thickness, orientation and phase function type and generate 4000 single-layer BSDFs and 12000 two-layer BSDFs as the training dataset. For each BSDF, we sample the $\omega_i$ and $\omega_o$ with $32\times32$ uniform stratified samples respectively. Starting from a sampled $\omega_i$ at the original location, Monte Carlo sampling in the media is performed in the media: sample a new position and sample a new direction with the phase function sampling. This process continues until the maximum depth (set as 20) is reached, or the ray exits the surface. We trace 100K samples for each $\omega_i$. We sample the directions with uniform sampling to guarantee sufficient samples at the grazing angles. Although the noise is still obvious, it does not affect the neural network training. We use 90\% of the dataset for training and the rest for validation.

\paragraph{Training.} The loss function is MAE (mean absolute error) of the difference between the ground truth and the single scattering computed from the output. Our network is implemented in the PyTorch framework; we also implement our single scattering model in PyTorch, making it automatically differentiable. We apply the Adam solver, where the learning rate is set as 0.001. The training samples are fed into the network in a mini-batch size of 32. For single-layer network, it took four hours on an NVIDIA 2080Ti GPU for training. A two-layer network (Fig.~\ref{fig:netTwoLayer}) took ten hours on an NVIDIA 2080Ti GPU for training. Note that the resulting networks are specific to a given number of layers, but general across layer parameters.





%% file: sec_results.tex
\section{Results and Comparison}
\label{sec:results}



We have implemented our algorithm inside the Mitsuba renderer~\shortcite{Mitsuba}. All timings in this section are measured on a 2.20GHz Intel i7 (40 threads) with 32 GB of memory. Since the evaluation of our model is analytic, it could also be easily implemented in any other rendering system and hardware, ray-traced or rasterized, CPU or GPU. We use mean squared error (MSE) to measure the image difference. The layered material settings are detailed in Table~\ref{tab:layerConfig}. 

Regarding the time cost of our model, for a single-layered material, our single-scattering has similar time cost to the microfacet model (given the similar form of Eqn.~\ref{eq:bsdf-flake}), and including multiple scattering doubles the cost. Multiple layers scale the cost correspondingly, as each layer will contribute a lobe. In this paper, we report the full rendering time, which includes both raytracing and shading

\input{paper_table_scene}

\myfigure{normalmap}{normalmap.png}{Normal mapping: we can use orientation mapping for the microflakes, instead of modifying the shading normal; this avoids the black artifacts of traditional normal mapping. We compare a microfacet surface to our method, where we use a single-layer surface SGGX microflake.}

\myfigure{sheen}{sheen.pdf}{Our model can be used to add a ``sheen'' or ``peach fuzz'' effect on top of any other BSDF, which approximates the effect of small fly-away fibers protruding out of the material. At low roughness, the common solution by Conty and Kulla~\shortcite{KullaConty:2017:revisiting} predicts very dark appearance. In contrast, our solution shows the desired sharp grazing effect even at low roughness.}

\mycfigure{fabric}{fabric.pdf}{For this fabric material, we use a two-layer model: the top layer is a fiber-like SGGX model with roughness 0.5, thickness as 1, with an orientation map, and the bottom map is a surface-like SGGX model with roughness 0.8, thickness as 5. The single scattering is computed with the analytical model, and the multiple scattering is computed using the analytic approximation with our network-predicted parameters. We compare our method against Guo et al.~\shortcite{Guo:2018:Layered} with equal time under the same material configuration and our result has much less noise. \revise{Note that our MSE is not much lower, due to the approximation of multiple scattering, which adds some numerical error but not noise. }}

\mycfigure{plant}{plant.png}{To model plant leaves, we use a three-layer configuration: two thin specular surface flake layers with roughness 0.05, enclosing a volumetric layer with a forward-scattering Henyey-Greenstein phase function. Orientation maps are used for the top and bottom layers and thickness map is used for the middle layer. Note the white specular reflection due to front lighting, and strongly peaked colored back-lighting due to forward scattering. The common alternative solution of using a Lambertian transmission lobe cannot achieve the peaked back-lighting effect, and remains much darker in transmission.}

\mycfigure{wood}{wood.png}{We apply SpongeCake to varnished wood with secondary specular highlights, using texture data from Marschner et al.~\shortcite{Marschner:2005:wood}. We use three layers: a coating layer with a surface-like SGGX model, a secondary specular layer with a fiber-like SGGX model with a textured fiber orientation and albedo, and a diffuse layer with a high-roughness surface-like SGGX flake. We compare our results against Guo et al.~\shortcite{Guo:2018:Layered} with equal time. Our method produces results with much less noise.}

\mycfigure{metal}{metal.png}{Previous work has focused on approximating multiple scattering in microfacet models (the effect is very prominent in e.g. rough metals). While this is not the main goal of our work, in practice we can obtain similar results by adding multiple scattering to a microflake layer. Our solution closely matches the reference for microflake multiple scattering.}

\paragraph{Model validation.} To validate the correctness of our single scattering model, we compare the rendered results using our single scattering model with a Monte Carlo simulation in Fig.~\ref{fig:MIS}. They produce identical results, which confirms the correctness of our single scattering derivation. We also test different sampling strategies: BSDF sampling, light sampling and their combination using multiple importance sampling (MIS). All of these sampling strategies provide identical results. Because our model supports MIS, it can be used in standard path tracing implementations.

In Fig.~\ref{fig:singleLayerMultiple}, \ref{fig:singleLayerMultiple_only}, \ref{fig:singleLayerLobe}, ~\ref{fig:twoLayerMultiple}, \ref{fig:twoLayerMultiple_only} and~\ref{fig:twoLayerLobe}, we compare the rendering results (single scattering + multiple scattering in Fig.~\ref{fig:singleLayerMultiple} and~\ref{fig:twoLayerMultiple} while multiple scattering only in Fig.~\ref{fig:singleLayerMultiple_only} and~\ref{fig:twoLayerMultiple_only}) and the lobes of our multiple scattering model against the ground truth (Monte Carlo simulated) on a set of single-layer and two-layer materials. More examples are shown in the supplemental material. We can see an overall good fit, and we discuss the more difficult cases in Section \ref{sec:limitations}. The Monte Carlo simulated approach for the ground truth is similar to Guo et al.~\shortcite{Guo:2018:Layered}, except no surface interfaces.


\paragraph{Similarity to microfacet model.} We first demonstrate that our model is able to produce appearance similar to that from microfacet models. This part is mostly a proof of concept: since microflakes and microfacets are different models with different parameter spaces, we do not aim at exact appearance match between the two models.


In Fig.~\ref{fig:compareMicrofacet}, we compare our model with the microfacet model on single-layered materials with varying roughness. Though we do not aim to produce an exact match of the appearance from the two models, it turns out that the difference is surprisingly minor, as shown in the difference images.

The good match indicates that there are connections between the microflake model and the microfacet model, which are also studied by Dupuy et al.~\shortcite{Dupuy:2016:unified}. Again, since our purpose is to demonstrate that our SpongeCake model can reproduce similar classic appearance (in addition to many more effects), we leave further study of the relationship between the models to the future.

\paragraph{Flake orientation mapping.} Normal mapping is a commonly used effect, which is known to cause artifacts due to areas with front-facing geometric but back-facing shading normal. One of the benefits of our model is that we can use orientation mapping for the microflakes, which is able to avoid the black artifacts naturally, without requiring any specific processing~\cite{Schussler:2017:normal}. We demonstrate this by using a normal (orientation) map in Fig.~\ref{fig:normalmap}. We compare a microfacet surface to our method, where we use a single-layer surface microflake with single scattering. The microfacet model produces black artifacts, while our method avoids this issue, since the half vector looking up the microflake distribution function in Eq.~\ref{eq:bsdf-flake} is not constrained to lie in a half-space and is valid for all orientations. While this does not make an attempt to simulate the underlying physics (light inter-reflection in deep grooves), it is a practical approach with plausible results. 

\paragraph{Fabric scene.} In this Fabric scene (Fig.~\ref{fig:fabric}) from Guo et al.~\shortcite{Guo:2018:Layered}, we use a two-layer model: the top layer is a fiber-like SGGX model with roughness 0.5, thickness as 1, with an orientation map, and the bottom map is a surface-like SGGX model with roughness 0.8, thickness as 5. We show the results with single scattering only, and full solutions. The single scattering is computed with the analytical model, and the multiple scattering is computed with the predicted parameters. We modified Guo et al.~\shortcite{Guo:2018:Layered} to handle layered materials without interfaces and use it to render a reference with higher sample rate. Our method produces very similar results to the reference. We also compare our method against Guo et al.~\shortcite{Guo:2018:Layered} with equal time. We find that our result has much less noise.

\paragraph{Plant scene.} In Fig.~\ref{fig:plant}, we use a three-layer model to model plant leaves: a thin surface flake layer with roughness 0.05 to represent front specular reflection, a volumetric layer with a forward-scattering Henyey-Greenstein phase function with $g = 0.7$ and another thin specular surface flake layer for the other side. Orientation maps are used for the top and bottom layers, and a thickness map is used for the middle layer. Note the distinct effects of both front lighting and back lighting: white specular reflection on the front, and strongly peaked colored back-lighting due to forward scattering. The common alternative solution of using a Lambertian transmission lobe is also demonstrated; the back-lighting is not anywhere as strong and peaked. This effect cannot be achieved without forward scattering, which is simple to model with our approach, but there is a lack of comparable analytic material models with this behavior. Existing solutions require a separate BSDF model just for this case~\cite{Burley:2015:pbs}.


\paragraph{Window shade scene.} In Fig.~\ref{teaser}, we model a layered fabric window shade material. Two SGGX models are used: the inside is a fiber-like flake with thickness 1 and roughness 0.1, and the outside layer has the same thickness and roughness 1.0 (essentially isotropic scattering). We show the indoor view with three lighting settings: sun only (left), lamp only (middle), and both sun and lamp (right). 

\paragraph{Wood scene.} In Fig.~\ref{fig:wood} we show three types of woods (padauk, walnut and curly maple), using texture data from Marschner et al.~\shortcite{Marschner:2005:wood}. The benefit of our model is that one does not need to implement an additional BRDF, in this case Marschner's wood model. We set up a layering with our model, where we use three layers: a coating layer with a surface-like SGGX model (roughness 0.05, and thickness 0.1), a secondary specular layer with a fiber-like SGGX model (roughness 0.2, thickness 1) with a textured fiber orientation and albedo, and a diffuse layer with a surface-like SGGX flake (roughness 0.8 and thickness 5, with textured albedo). We do not use a substrate BRDF in this example. We compare our method against Guo et al.~\shortcite{Guo:2018:Layered} with equal time. By comparison, we find that our method has much less noise and the MSE is much smaller.

\myfigure{whiteFurnace}{whiteFurnace.pdf}{A white furnace test for a SpongeCake two-layered material with no absorption. Our delta transmission and single scattering results in darker pixel values as expected. Our multiple scattering closely approximates a constant image, despite some inaccuracies due to being an approximation. Using ground-truth Monte Carlo multiple scattering will produce a constant image, modulo integration error.}

\paragraph{Sheen.} Our model can also be used to add a ``sheen'' or ``peach fuzz'' effect on top of any other BSDF, which approximates the effect of small fly-away fibers protruding out of the material. A common solution for this effect is the BRDF introduced by Conty and Kulla~\shortcite{KullaConty:2017:revisiting}, which is a microfacet-based layer with a fiber-like instead of surface-like normal distribution function. While this is a useful model and our method is not designed to replace it, we offer a few benefits. First, our solution (when limited to single scattering) is completely analytic and does not require numerical fits to the shadowing-masking behavior. Second, at low roughness, Conty and Kulla's model becomes very dark due to microfacet shadowing/masking becoming prominent and virtually eliminating the reflection itself; in contrast, our solution shows the desired sharp grazing effect even at low roughness (Fig.~\ref{fig:sheen}).

\paragraph{Multiple scattering for rough metals.} Missing energy is a common issue for microfacet models due to ignored multiple scattering, and several research efforts have focused on addressing this. As shown in Fig.~\ref{fig:metal}, our method with single scattering achieves similar appearance to the microfacet model for a rough copper material. With our multiple scattering added, our method is able to provide multiple scattered results that closely resemble the reference.

\paragraph{White furnace test.} The SpongeCake model is energy-conserving when using ground-truth evaluation; however, our analytic multiple scattering is an approximation to the true multiple scattering. Therefore, if we model a material with no absorption and render it in a white furnace (i.e. a constant environment with unit incoming radiance), we should expect our single scattering results to always pass the test with pixel values at most one. This is exactly what we observe in the experiments (Fig.~\ref{fig:whiteFurnace}). Adding Monte Carlo multiple scattering will produce a constant image modulo integration error, while our multiple scattering (due to being an approximation) will lead to an almost constant image with some loss of accuracy. We have not observed problems due to this in any rendered results.


\section{Discussion and Limitations} \label{sec:limitations}

\paragraph{Refraction.} An obvious limitation of our model is that volumetric layers cannot simulate refraction (though we could still add volumetric layers on top of a standard refractive BSDF substrate). Note that our method can still reasonably model thin flat slabs of a refractive material (e.g. glass panes); as long as the two interfaces are parallel, the resulting ray direction is unaffected. If different normal mapping is applied to the two interfaces, such as in thin jade slab examples shown in Guo et al.'s paper, our model is not capable of approximating the effect; however, this is arguably not a common use case.

\myfigure{limitation}{limitation_fiber.png}{Our multiple scattering approximation does not provide perfect fitting in some cases, e.g. a fiber-like flake with low roughness. This is not a flaw of the neural network, but instead a limitation of the expressiveness of the two added lobes in our approximation. However, the inaccuracy is less obvious in final rendered results, since the proportion of multiple scattering is also low compared to single scattering in this case. The first two columns represent the entire BSDF (top and bottom hemispheres), with pixel rows corresponding to a discretization of incoming directions, and pixel columns corresponding to outgoing directions. The latter four columns visualize the outgoing lobe given a fixed incoming direction at the specified angle {(radians)} $\theta$ with pixel rows corresponding to the elevation angle of the outgoing direction, and pixel columns corresponding to the azimuth angle of the outgoing directions.}


\paragraph{Orientation mapping, other media types.} Another limitation is that non-microflake layers cannot be easily orientation-mapped (such as the HG layer in our plant example). A solution to this may be to extend the microflake types supported from mirror reflectance to Lambertian or rough reflection and transmission. There is no known analytic solution for the phase functions of such microflakes, though an implementation based on table look-ups may be sufficient. Also, we currently do not support depth-varying densities or non-exponential (correlated) media, though we believe that Eqs.~\ref{eq:bsdf1} and~\ref{eq:bsdf3} may be extended to include these effects with some additional effort.

\paragraph{Expressive power of added lobes.} One limitation is that our multiple scattering network may not always be able to find accurate fits. A typical failure case is on fiber microflakes with low roughness, as shown in Fig.~\ref{fig:limitation}. However, this is not the fault of the network; there are no existing single scattering parameters that would give a distribution matching the multiple scattering distribution. The two diagonals in the figure correspond to the case where the half vector is aligned with the fiber orientation. For such a configuration, the fiber-like phase function reflects low energy. The multiple scattering does not have black diagonals due to the ``blurring'' of this effect with more scattering events. Our method approximates multiple scattering with single scattering lobes of modified parameters, and cannot find a lobe that exactly produces the desired behavior. One might expect that higher roughness single scattering lobes could approximate the effect better, but in fact our method tends to find parameters with lower roughness, since they lead to smaller error overall. Fortunately, when rendered with both single and multiple scattering, a natural balance helps us to reduce the difference to the minimum. That is, when the roughness of fibers is low, our predicted multiple scattering is less accurate, but the proportion of multiple scattered energy is also low. When the fibers are of high roughness, multiple scattering dominates, but our network is also able to perform good parameter fits.

\paragraph{\revise{Spatially varying materials.}}\revise{Our model is able to support spatially-varying materials. If the parameters are known before rendering, the network inference happens during precomputation. With procedurally evaluated materials at render time, our model would however need to run the neural network inference during rendering.}


\paragraph{More discussion.}
Note that we are currently only concerned with BSDF models, thus we ignore BSSRDF effects (light entering and exiting from different spatial locations on the top layer interface). This limitation is shared with all existing layered BSDF work, but considering BSSRDFs could be an interesting future extension.

Being a useful forward model, SpongeCake may also be helpful when used for inverse rendering\revise{, even for inverse rendering of layered materials~\cite{Bati:2021:layered}}, since the model is analytic and has good expressive power for representation with a relatively small parameter space, and is also easily made differentiable. For example, in Fig~\ref{fig:metal}, we manually set the parameters to match our results with those from microfacet models. This process can be automated with the help of differentiable rendering and gradient-descent based optimization methods.

Finally, while our model can create a wide range of appearances, the techniques to achieve them are not standard knowledge among artists, and may require either a steep learning curve or an artist-friendly reinterpretation.






%% file: paper_table_scene.tex
 \begin{table}[htbp]
  \caption{\label{tab:layerConfig}The parameter settings used in our scenes. The main parameters for each layer include albedo $\gamma$, roughness $\alpha$ (for hg, $\alpha$ is the g),  Schlick reflectance $f_0$, thickness $T$, the phase function type $k$ (fiber or surface microflake or hg) and orientation $\omega_p$. Here, $\alpha$ for a media with hg phase function represents the mean cosine of scattering interactions. The microflake reflectance is controlled by a constant \emph{albedo} and a Schlick Fresnel approximation with $f_0$ controlling the reflectance at 0 degrees (so $f_0 = (1,1,1)$ means no Fresnel term applied). }
\centering
\setlength{\tabcolsep}{2pt}
\begin{tabular}{lcccccccc}
\toprule
      \multicolumn{1}{c}{Scene}
	 & \multicolumn{1}{c}{Layer}
	 & \multicolumn{5}{c}{Parameters}\\	
		\cmidrule(lr){3-8}
	 & 
	 & \multicolumn{1}{c}{ $k$}
	 & \multicolumn{1}{c}{ $\gamma$}
	 & \multicolumn{1}{c}{ $\alpha$}
	 & \multicolumn{1}{c}{ $f_0$}
	 & \multicolumn{1}{c}{ $T$}
	 & \multicolumn{1}{c}{ $\omega_p$}\\	
		\midrule
		Window & L 1 & fiber &(0.9,0.9,0.7) &0.1&(1,1,1)&1.0 & map\\ 	\cmidrule(lr){2-8}
		Shade & L 2 &fiber & (0.9,0.9,0.7) &1.0&(1,1,1)&1.0 & map\\ \hline 
		  & L 1 &fiber  &(0.7,0.1,0.1) &0.1&(1,1,1)&1.0 & map\\
		\cmidrule(lr){2-8}
		Fabric  & L 2 &surface  &(0.7,0.1,0.1) &0.8&(1,1,1)&5.0 & (0,0,1)\\ 
		\hline 
		  & L 1 &surface  &(1,1,1) &0.05&(0.1,0.1,0.1)&0.05& map\\ 
		\cmidrule(lr){2-8}
		Plant  & L 2 & hg  &(0.7,0.1,0.1) &0.7 &(1,1,1) & map & --\\ 
		\cmidrule(lr){2-8} 
		  & L 3 &surface  &(1,1,1) &0.05&(1,1,1)&0.002& map\\ 
		\hline 
		  & L 1 &surface  & (1,1,1) &0.05&(0.1,0.1,0.1)&0.1& (0,0,1)\\ 
		\cmidrule(lr){2-8}
		Wood  & L 2 & fiber  & map & 0.2 &(1,1,1) & 1.0 & map \\ 
		\cmidrule(lr){2-8}
		  & L 3 &surface  &(1,1,1) &0.8 &(1,1,1)&5& (0,0,1)\\
\bottomrule

\end{tabular}

\end{table}

%% file: sec_conclusion.tex
\section{Conclusion and Future Work}
\label{sec:conclusion}

We have presented SpongeCake, a layered BSDF model where each layer is a volumetric scattering medium, and without any reflecting and refracting interfaces between the layers (except an optional substrate). With our model, we have derived an exact analytic solution for single scattering, regardless of the number of volumetric layers. We have approximated the multiple scattering by adding two new lobes (an additional single scattering lobe and a Lambertian lobe), whose parameters are estimated by a parameter mapping neural network. Thanks to the analytic formulation, our model is very fast to evaluate and sample, and does not introduce noise to the BSDF evaluation. Through various parameter settings, we have demonstrated that our model handles many types of materials, like plastics, wood, cloth, and leaves, as well as supporting orientation mapping, anisotropy, and fiber sheen effects with no additional effort. Interactive and real-time applications of the SpongeCake model should be possible in the near future.

In the future, it would also be interesting (and challenging) to explore possibilities to unify surface and media layers further. Depth-varying densities, diffuse or one-sided microflakes, and other extensions may be useful to achieve more effects and deeper understanding. Extending our model to deal with correlated participating media~\cite{Jarabo:2018:Correlated, dEon:2018:Nonexponential} to study how non-exponential falloff would expand the range of surface appearance could also be interesting, as would the incorporation of non-local BSSRDF effects. Since our SpongeCake model demonstrates the ability to represent a large variety of appearance, it could also potentially act as a useful differentiable forward model to fit physical materials from real measurements, or help with inverse rendering problems.

\begin{acks}
We thank the reviewers for the valuable comments. This work has been partially supported by the National Natural Science Foundation of China under grant No. 62172220. Ling-Qi Yan is supported by gift funds from Adobe, Dimension 5 and XVerse.
\end{acks}